\documentclass[aps,pra,superscriptaddress,twocolumn,showpacs,floatfix]{revtex4-2}

\usepackage{amssymb,amsmath,amsfonts,mathrsfs,bm, bbm, graphicx, epsfig, epstopdf}
\usepackage[version=3]{mhchem}
\usepackage{braket}
\usepackage{hyperref}
\usepackage{color}
\usepackage{xcolor}
\usepackage{adjustbox}
\usepackage{csquotes}
\usepackage{float}
\usepackage{colortbl}
\usepackage{siunitx}
\DeclareSIUnit{\nothing}{\relax}
\usepackage{makecell}
\usepackage[T1]{fontenc}
\usepackage{soul}

\date{\today}
\begin{document}
\title{Machine-Learning-Enhanced Quantum Optical Storage in Solids}

\author{Yisheng Lei}
\affiliation{Department of Electrical and Computer Engineering and Applied Physics Program, Northwestern University, Evanston, IL 60208, USA}

\author{Haechan An}
\affiliation{Department of Electrical and Computer Engineering and Applied Physics Program, Northwestern University, Evanston, IL 60208, USA}
\affiliation{Elmore Family School of Electrical and Computer Engineering, Purdue University, West Lafayette, Indiana 47907, USA}

\author{Zongfeng Li}
\affiliation{Department of Electrical and Computer Engineering and Applied Physics Program, Northwestern University, Evanston, IL 60208, USA}

\author{Mahdi Hosseini}
\thanks{Corresponding author}
\email{mh@northwestern.edu}
\affiliation{Department of Electrical and Computer Engineering and Applied Physics Program, Northwestern University, Evanston, IL 60208, USA}
\affiliation{Elmore Family School of Electrical and Computer Engineering, Purdue University, West Lafayette, Indiana 47907, USA}

\begin{abstract}
Quantum memory devices with high storage efficiency and bandwidth are essential elements for future quantum networks. Solid-state quantum memories can provide broadband storage, but they primarily suffer from low storage efficiency. We use passive optimization and machine learning techniques to demonstrate nearly a 6-fold enhancement in quantum memory efficiency. In this regime, we demonstrate coherent and single-photon-level storage with a high signal-to-noise ratio. The optimization technique presented here can be applied to most solid-state quantum memories to significantly improve the storage efficiency without compromising the memory bandwidth. 
\end{abstract}

\maketitle{}
Rare-earth-ion doped solids have been attractive platforms for the development of quantum optical memories \cite{lei2023quantum}. The atomic frequency comb technique \cite{afzelius2009multimode} has become the primary storage protocol in solids due to its broadband and low-noise properties. Typically, a 2-pulse train pumping sequence is used to perform spectral tailoring and create an atomic frequency comb (AFC) \cite{chaneliere2010efficient, jiang2023quantum}. Improving optical depth using impedance-matched resonators can be used to improve storage efficiency at the expense of lowering the memory bandwidth \cite{afzelius2010impedance, moiseev2010efficient}. Optimizing the pumping and preparation sequence is also crucial for better spectral tailoring leading to higher storage efficiency. In the case of laser-cooled atoms, machine-learning optimization has been deployed to enhance atom trapping and cooling \cite{tranter2018multiparameter, vendeiro2022machine}. Such optimization has not been explored in the context of solid-state quantum memories.
In this article, we perform machine-learning optimization of AFC quantum storage in a Tm$^{3+}$: YAG crystal. Tm$^{3+}$ ions in solids have optical transition wavelengths close to those of Rubidium atoms, making them good candidates for building hybrid quantum networks \cite{gu2024hybrid}. Many experiments are performed using Tm$^{3+}$: YAG Crystal \cite{chaneliere2010efficient, davidson2020improved}, as well as Tm$^{3+}$: YGG Crystal \cite{thiel2014tm, askarani2021long}, Tm$^{3+}$: LiNbO$_\text{3}$ crystal \cite{sinclair2014spectral, askarani2020entanglement}, and Tm$^{3+}$ ions doped in Lithium Niobate on Insulator \cite{dutta2023atomic}. Tm$^{3+}$ ions in YAG crystal have a long optical coherence time of 100$\mu$s, a ground state lifetime of over 1s at around 1K, and a high branching ratio of 25\% \cite{louchet2007branching}, making it a good system for spectral tailoring for photon storage.
Our experiment is carried out using a tabletop cryostat at 3.5K and a magnetic field produced by a compact permanent magnet. We first passively enhance the optical depth without compromising bandwidth by routing the laser beam multiple times through the crystal. We then run a genetic algorithm to design a more efficient spectral preparation sequence. We show that the combination of these techniques can lead to a significant improvement in storage efficiency. We also demonstrate coherent and single-photon-level storage with a high signal-to-noise ratio.
\begin{figure}[!h]
\centerline{\includegraphics[width=0.85\columnwidth]{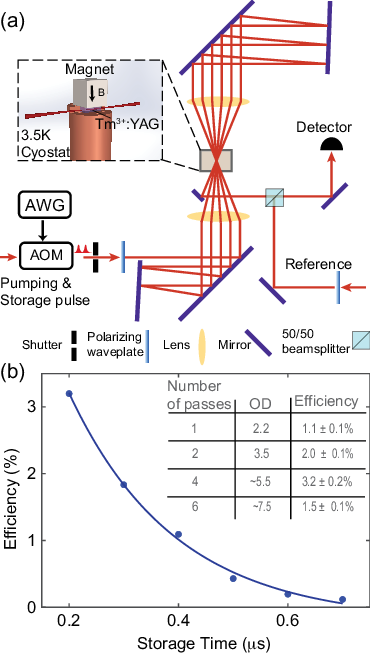}}
	\caption{(a) The experimental setup is shown where a Tm$^{3+}$: YAG crystal is placed inside a 3.5K cryostat below a permanent magnet. A few mirrors (M) are used to direct the optical signal to pass multiple times through the crystal. For interferometric detection, the signal is interfered with a reference light with 80Hz frequency difference. AOM: acousto-optic modulator, AWG: arbitrary waveform generator. (b) Results of multi-pass 2-pulse train atomic frequency comb (AFC) storage. Four-pass echo efficiency is plotted as a function of the time separations between the two pulses used to create AFC. An exponential fit with a time constant of 194~ns indicates the storage time. The inset shows the echo efficiency and the estimated optical depth for different numbers of passes through the memory with a storage time of 200~ns.}
	\label{Fig1}
\end{figure}

The experimental setup is shown in Fig.~\ref{Fig1} (a). The Tm$^{3+}$: YAG crystal with dimensions $4\times5\times10$ mm (crystal axis $\langle$001$\rangle$, $\langle$1$\Bar{1}$0$\rangle$, $\langle$110$\rangle$) has a doping concentration of 0.1\%. The laser beam (Toptica DL Pro) propagates parallel to the $\langle$110$\rangle$ axis with linear polarization (E $\parallel$ $\langle$1$\Bar{1}$0$\rangle$ axis). The optical transition $^{\text{3}}$H${\text{4}}$ $\leftrightarrow$ $^{\text{3}}$H${\text{6}}$ occurs at a wavelength of $\lambda=$793.373 nm. A permanent magnet produces a magnetic field of 600~G along the $\langle$001$\rangle$ axis, lifting the degeneracy of Tm$^{3+}$ ions with a nuclear spin of $1/2$. One of the two ground states can serve as a shelving state for spectral preparation. With spectral hole burning, its hole and side-holes (due to inhomogeneous broadening) are separated by 3.75 MHz, indicating that the excited state is split by 3.75 MHz, and the ground state splitting is determined to be 17.3 MHz. The two-pulse echo technique is used to measure the homogeneous linewidth. The echo intensities are measured with different time separations between the two pulses, where an exponential fitting outputs a maximum coherence time of 60 $\mu$s, which is inverse of the homogeneous linewidth. Additional measurements of spectral hole decay with double-exponential fitting output a fast decay lifetime of 10~ms and a slow decay lifetime of 110~ms, corresponding to the intermediate level $^{\text{3}}$F$_{\text{4}}$ lifetime and the ground state lifetime \cite{sinclair2021optical}.

We implement an AFC protocol for storage, which relies on spectral tailoring and atomic rephasing \cite{afzelius2009multimode}. A common approach for spectral tailoring is to use a two-pulse train to optically pump atoms into an AFC. We first implement AFC using two identical Gaussian pulses of width 50 ns separated by $t_s=200$~ns with a 5 $\mu$s delay between pulse pairs, repeated 15,000 times followed by a 20 ms wait time. A 50 ns weak coherent pulse is then sent to carry out storage. This approach gives rise to a memory bandwidth of 20 MHz, a storage time of $t_s=200$~ns, and a memory efficiency of 1.1~$\pm$ 0.1~\%.

The AFC memory efficiency ideally scales with optical depth(OD) as \cite{lauritzen2011approaches}:
\begin{equation}\label{Equ1}
\eta = \frac{d^2}{F^2}\text{exp}(-\frac{d}{F})\text{exp}(-\frac{1}{F^2}\frac{\pi^2}{4ln2})\text{exp}(-d_0), 
\end{equation}
where $d$ is the OD of the atomic medium, $F$ is the finesse of the comb, and $d_0$ is the total background absorption. Finesse is defined as the ratio between the spacing and width of the Gaussian comb lines. The typical finesse of a frequency comb produced by a two-pulse-train pump is around 2-3. In the case of an ideal Gaussian-teeth AFC, the maximum storage efficiency can be achieved with an OD of 4-6. Increasing the optical depth can be achieved simply by using a longer crystal, but it would require more cooling power and a larger sample space in the cryostat. In our experiment, we first route the laser beam to pass through the crystal multiple times. For different numbers of passes, both echo efficiency and OD are measured (see Fig. \ref{Fig1} (b)). Maximum efficiency was observed for the four-pass configuration. At higher numbers of passes, storage efficiency drops because the pumping laser intensity and time are limited by laser heating and ground-state lifetime, respectively. We also observe that the OD does not show a linear relationship with the number of passes, which we attribute to the non-zero overlap between different beam paths. 

\begin{figure*}[!ht]
	\centerline{\includegraphics[width=1.5\columnwidth, angle=0]{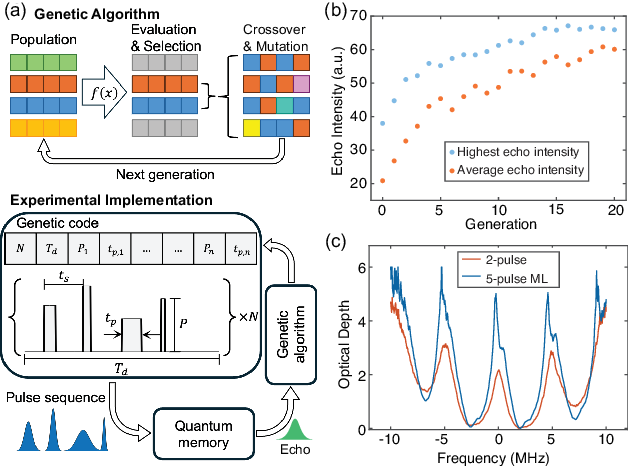}}
	\caption{(a) Flow diagram of the genetic algorithm (top) and its implementation in atomic frequency comb memory (bottom). The genetic algorithm evaluates parameter sets and generates the next parameter set from the current parameter sets and their evaluation results. Repetition of this evolution of parameter sets leads to sets with good evaluation results. In the context of atomic frequency comb memory, the parameters to be optimized are the design parameters of the pumping pulse sequence, while the evaluation and results focus on the efficiency of the atomic frequency comb memory and the echo. (b) The mean and best value of echo intensity for every generation of the machine learning algorithm is plotted. (c) Optical depth after pumping is plotted as a function of relative frequency for the traditional 2-pulse-train pumping sequence and a 5-pulse machine-learning sequence.}
	\label{Fig2}
\end{figure*}
The efficiency relationship in Eq.~\ref{Equ1} assumes a simple atomic level structure and an infinite comb with perfect Gaussian teeth. In practice, the experimental effects of optoelectronic devices, optical and electronic delays, and the complex atomic-level structure can cause deviations from predictions obtained through a straightforward Fourier transform of the ideal pumping pulse train in the time domain. Machine learning optimization can offer significant benefits in addressing these complexities. Thus, to further improve the pumping and storage efficiency for the four-pass configuration, we employ machine learning based on a genetic algorithm. Genetic algorithms and their variants are recognized as derivative-free optimization algorithms capable of delivering high-quality solutions even with straightforward implementation. This makes genetic algorithms attractive optimization tools in various fields, including atomic and optical experiments \cite{judson1992teaching,hornung2000optimal,chen2021genetic,robertson2024machine}. The genetic algorithm is inspired by the theory of natural evolution, and its flow is schematically shown in Fig.~\ref{Fig2} (a). Individuals have a genetic code consisting of parameters to be optimized. Each individual is evaluated on how close it is to the optimization goal, and individuals with good evaluation results (fitness) are selected to create the next generation. The genetic codes of offspring are made by crossing over and mutating the parents’ genetic code. Keeping the individuals with the best fitness (elites) to the next generation can accelerate the optimization. Repeating this process for several generations leads to the evolution of individuals and delivers solutions to an optimization problem. We send the same 50 ns pulse as an input signal to AFC memory, use the first echo intensity as the fitness function, and employ the genetic algorithm to optimize pulse parameters as shown in Fig.~\ref{Fig2} (a). We note that employing echo intensity, or equivalently, memory efficiency, as the cost function, ensures optimal memory performance for a specific memory bandwidth (determined by input pulse bandwidth) at a given storage time.
We fist consider a two-pulse pumping sequence and set the pulse widths ($t_p=~$20-100ns), amplitudes ($P=$0 - 0.6V, corresponding to approximately 0-10mW) laser output, loop duration ($T_d=$2 - 10$\mu$s), and the number of loops ($N=$5k - 50k) as constrained variables, while keeping pulse separations fixed at $t_s=$200ns. We set both the total number of generations and population to 20. Parents are selected through tournament selection with a tournament size of 10. Three elites are kept to the next generation. 

Fig.~\ref{Fig2} (b) shows the best and mean fitness of each generation, with fitness improving over generations. The small fluctuation in the best fitness is due to the fluctuation of the echo intensity for each measurement. The fluctuation in the experimental result directly affects optimization and becomes the resolution of optimization. We reduce fluctuation by averaging 25 echo measurements, and apply Gaussian fitting to find echo intensity as a measure of efficiency. We then consider n-pulse train sequences with $n=1-10$ and identify the five-pulse sequence to be the most optimum with a memory efficiency of 6.4$\pm$0.1\%, which is a 100\% increase compared to simple two-pulse-train pumping. The optimal pulse parameters are found to be $P,t_p=$(0.11 V, 21 ns; 0.10 V, 55 ns; 0.31 V, 60 ns; 0.51 V, 67 ns; 0.10 V, 62 ns), with a loop duration $T_d=$4.2 $\mu$s and $N=$45, 600 loops. As seen in Fig.~\ref{Fig2} (c), the optimized pumping sequence results in less background absorption and a higher optical depth. The comb finesse $F$ = 3.68, optical depth $d \sim$ 5, and optical background absorption $d_0 \sim$ 1.5, which should result in a theoretical efficiency of 8.1\% based on Eq. \ref{Equ1}. We attribute this slight discrepancy to the imperfect comb shape and finite comb length. We also tried more complicated sequences, such as multi-stage pumping (running a few different pulse sequences sequentially), but it didn’t yield better results.

We also carry out interferometric measurements and evaluate the machine learning performance using the coherent echo amplitude as the fitness. We interfere the weak storage pulse with a reference light of 80 MHz frequency difference on a 50:50 beamsplitter. The amplitude of the interference envelope is set as the fitness, and the machine learning outputs a pumping sequence with a memory efficiency increase of 80\% compared to the simple 2-pulse pumping.

\begin{figure}[!t]
\centerline{\includegraphics[width=0.85\columnwidth, angle=0]{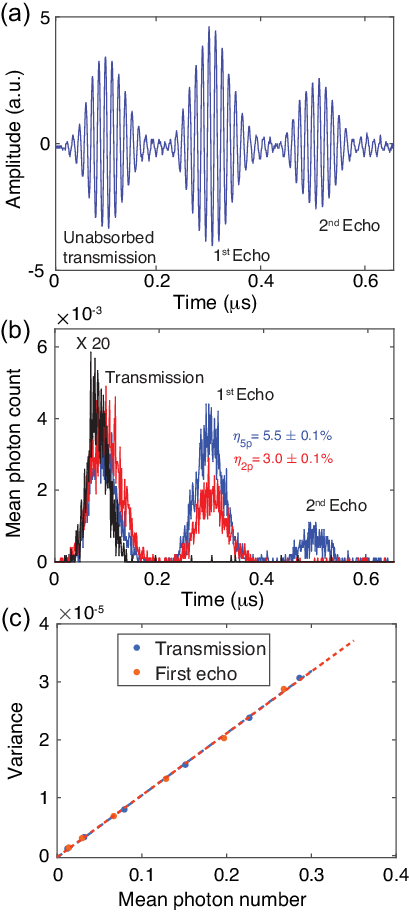}}
	\caption{(a) The interferometric measurement of memory output for the optimized five-pulse sequence is plotted, indicating the coherent nature of the storage. (b) Single-photon level storage is presented where the black curve represents the input storage pulse measured by tuning the laser frequency off-resonant, while the signal in blue and red represents the memory outgoing signal obtained from the five-pulse machine learning and the two-pulse-train pumping sequence, respectively. (c) The variance of the photon number for the first echo and leakage pulses is plotted against the average photon number of the first echo, obtained from $10^4$ storage events. Linear fittings (blue and red dashed lines) output slopes of 1.073/$10^4$ and 1.068/$10^4$ for transmission and first echo pulses, respectively, in agreement with the expected shot-noise limit.}
	\label{Fig3}
\end{figure}

We also perform single-photon level storage using an attenuated coherent pulse and a mechanical shutter to fully block the pumping beam during storage. After each spectral preparation and 20 ms waiting, 100 weak storage pulses separated by 10 $\mu$s are sent to the crystal. A semiconductor avalanche single photon detector with an efficiency of 60\% (Laser Components, Single Photon Counting Module) is turned on for 2 $\mu$s for every storage pulse to record its transmission and echo pulses. In total, 10,000 storage events are performed and measured. Single photon measurements with five different average input photon numbers have been performed. Average photon numbers of the first echo range from 0.02 to 0.3. The input storage pulse intensities are measured by tuning the laser frequency off-resonant and adding a neutral density filter to reduce its intensity to 5\%, as shown in Fig. \ref{Fig3} (b). A memory efficiency of 5.5$\pm$0.1\% for the 5-pulse machine learning pumping sequence is measured, and 3.0$\pm$0.1\% for the simple 2-pulse sequence. In the case of 2-pulse-train pumping, the resulting less-than-ideal AFC structure leads to dispersion \cite{bonarota2012atomic} of the leakage light (unabsorbed pulse, depicted in red). By measuring the variance of single-photon detection events (Fig. \ref{Fig3} (c)), we confirm that the noise of the echo signal is shot-noise limited, i.e., $Var(n_{ph})\simeq\langle n_{ph}\rangle$, complementing the coherent storage results from interferometric measurement.

To further improve efficiency, moderate modifications are needed to increase the finesse of the AFC and reduce the background absorption. As shown previously, this can be achieved by reducing the laser linewidth and crystal temperature \cite{jobez2016towards}. In our experiment, the laser linewidth and the ground state lifetime of Tm$^{3+}$ ions limit the performance of spectral tailoring. Frequency locking of the laser can reduce the spectral linewidth, and lower temperature can prolong the ground state lifetime. Further reducing the cryostat's vibrations will help to run more sophisticated genetic algorithms with faster and more accurate optimizations. Combining our proof-of-concept optimization with these improvements, the comb finesse $F$ = 4 - 5, optical depth $d \sim$ 8 (in a 6-pass configuration), and optical background absorption $d_0 \sim$ 0.1 can be achieved, resulting in an efficiency of $\sim$40\% without compromising storage bandwidth. This is already very close to the maximum AFC efficiency expected in the case of forward retrieval, which is 54\% \cite{lauritzen2011approaches}. Cavity impedance matching conditions are another way to increase memory efficiency \cite{davidson2020improved}, albeit at the expense of lowering the memory bandwidth. Narrow laser spectral linewidth will also help to prepare a comb with smaller $\Delta$, which will give a longer storage time. Other pumping sequences such as intrinsic pumping method \cite{askarani2021long} can be employed to enhance the memory bandwidths up to $\sim$10 GHz.

In conclusion, we demonstrate that by optimizing effective OD via multi-pass configuration and employing machine-learning-assisted memory preparation, the efficiency of quantum optical storage can increase by about a factor of 6 in a Tm$^{3+}$: YAG crystal. We conduct classical, coherent, and single-photon-level storage to compare optimization results and characterize quantum noise. Our findings indicate that classical machine learning can significantly enhance quantum storage performance without introducing intensity or phase noise to the system. This method is applicable to other storage protocols developed in various rare-earth solids.

 We acknowledge the support from National Science Foundation Award number 2410054 and U.S. Department of Energy, Office of Science, Office of Advanced Scientific Computing Research, through the Quantum Internet to Accelerate Scientific Discovery Program under Field Work Proposal 3ERKJ381.

The authors declare no conflicts of interest.\\

Data underlying the results presented in this paper are not publicly available at this time but may be obtained from the authors upon reasonable request.

\bibliography{sample}{}

\end{document}